# Beyond Sequential Prediction: Learning Financial Market Dynamics in Volatile and Non-Stationary Environments through Sentiment-Conditioned Generative Modelling

Alexis Lazanas*

Department of Mechanical Engineering and Aeronautics,
University of Patras, Rion-Patras, Greece
Email: alexlas@upatras.gr

Spyridon Karpouzis

FICC Market Risk, BNP Paribas CIB,
London, United Kingdom
Email: spyros.karpouzis@uk.bnpparibas.com

*Correspondent author

**Abstract:** The problem of time-series forecasting in non-stationary and complex environments is a challenging task in machine learning, especially with heterogeneous numerical and textual data present. Traditional statistical models like AutoRegressive Integrated Moving Average (ARIMA) are based on the assumptions of linearity and stationarity, whereas recurrent neural networks like Long Short-Term Memory (LSTM) models do not necessarily represent distributional properties in highly volatile settings. This paper proposes a hybrid model that combines Generative Adversarial Networks (GANs) with Natural Language Processing (NLP)-based sentiment analysis to enable sentiment-conditioned time-series prediction. The model integrates adversarial learning on numerical sequences with contextual sentiment representations derived from unstructured text, enabling them to be jointly modelled to capture temporal dynamics and exogenous information. These results demonstrate the promise of hybrid generative and language-aware methods to enhance prediction robustness in non-stationary environments.

**Biographical notes:** Alexis Lazanas (BSc, PhD) holds a Laboratory Lecturer position at the Department of Mechanical Engineering and Aeronautics, University of Patras, Greece. Currently, he teaches Computer Programming, IS Analysis and Design, and Computer-Simulation Modelling. His research interests focus on Recommender Systems, Machine/Deep Learning, NLP, NNs, LLMs, Intelligent DSS, and Knowledge-driven frameworks for applying AI to organisational decision-making in complex socio-technical systems.

Spyridon Karpouzis (BEng, MSc) studied Mechanical Engineering and Aeronautics at the University of Patras, Greece. In 2025, he graduated from Imperial College Business School with a master's degree in Risk Management and Financial Engineering. He is currently employed in the Corporate Investment Banking division at BNP in London. His main interests are Machine Learning models, Quantitative Research, and Financial Market Risk Analysis.

## 1 Introduction

Time-series forecasting in non-stationary and noisy settings has been a key problem in machine learning, especially when the underlying dynamics are nonlinear and the information sources are heterogeneous. Classical methods of statistics, including ARIMA models, are based on restrictive assumptions of linearity and stationarity that limit their ability to capture complex temporal relationships in real-world data (Billings and Yang, 2006; Engle, 1982). Although these models remain convenient benchmarks, they tend to be overly limited in their representational ability in contemporary data landscapes that are volatile, regime-altering, and latent in their impact.

The emergence of deep learning has greatly contributed to time-series modelling by enabling data-driven representation learning without strong distributional assumptions. LSTM-based recurrent neural networks (RNNs) have gained popularity for their ability to learn long-term temporal dependencies (Hochreiter and Schmidhuber, 1997; Bengio et al., 1994). Models based on LSTMs have been shown to outperform classical statistical models in various sequence prediction problems, such as financial time-series prediction (Fischer and Krauss, 2018; Nelson et al., 2017). However, LSTM models are generally trained in a purely discriminative fashion and may not learn distributional properties well in highly volatile or regime-switching settings.

Another paradigm of learning is generative modelling, which, instead of making point predictions, explicitly models the underlying data distribution. GANs introduce an adversarial training process that has proven successful for training complex, high-dimensional data distributions (Goodfellow et al., 2014). Recent works have shown that GAN-based architectures possess the potential to be used sequentially and in time series, as well as in applications with noisy and non-stationary data (Zhang et al., 2019; Vuletić et al., 2023; Brophy et al., 2023). These results indicate that adversarial learning can improve robustness and generalisation in difficult prediction environments.

Simultaneously, the growing access to unstructured textual information has spurred the adoption of Natural Language Processing (NLP) methods in predictive modelling pipelines. Contextual information that cannot be encoded in numerical time-series can be encoded in textual data obtained in news articles, online and social media platforms. Other applications of NLP, such as sentiment analysis, make it possible to quantify subjective data, providing greater explanatory power in areas where human perception and collective action determine the dynamics (Bollen et al., 2011; Derakhshan and Beigy, 2019; Puh and Bagić Babac, 2023; Tuan et al., 2024). Although increasing attention has been paid to integrating textual and numerical data, current methods tend to treat these streams of information separately or apply superficial fusion techniques.

Regardless of these developments, current forecasting methods have significant shortcomings when applied to a non-stationary and volatile financial context. Statistical models can provide interpretable baselines but cannot capture nonlinear dependencies and sudden regime shifts. Deep recurrent models enhance temporal representations but tend to be discriminative predictors and may fail to adequately represent the conditional distribution of future observations when uncertainty is present. Furthermore, numerous sentiment-sensitive forecasting methods rely on shallow fusion, in which textual and numerical data are merely fused without specifying how to condition the learning. Such problems inspire the creation of a hybrid generative model that combines adversarial sequence learning with sentiment-based contextual information.

Based on the above, the research question of the current paper is to conduct a comparative analysis of various modelling strategies for financial time-series prediction in non-stationary and volatile environments, with the intent of identifying a framework that better incorporates numerical market data and sentiment-based textual information. The principal contributions of this research may be summarised in the following way:

- A comparative analysis of statistical (ARIMA), deep learning (LSTM), and generative (GAN-based) models of financial time-series forecasting in non-stationary conditions.
- The creation of a sentiment-conditioned GAN system, which combines numerical market data with textual sentiment data in a single predictive system.
- An empirical investigation illustrating the performance of various modelling paradigms in an asset-dependent manner, with the benefits of hybrid generative strategies in volatile and sentiment-driven markets.

In a broader sense, the present study is consistent with earlier research on smart, data-driven modelling frameworks that incorporate heterogeneous sources of information into predictive systems (Dedotsi et al., 2023), while also drawing on methodological insights from hybrid generative and language-aware models.

The rest of the paper is structured as follows: Section 2 reviews the related literature on time-series forecasting systems, generative adversarial models, and NLP-based systems. Section 3 Section 3 presents the problem formulation and the proposed GAN–NLP framework. Section 4 describes the experimental setup, data sources, and evaluation protocol. Section 5 reports the empirical results and comparative analysis. Finally, Section 6 discusses the findings, limitations, and directions for future research.

## 2 Related Work

Time-series prediction research has evolved beyond traditional statistical modelling to data-driven machine learning and deep learning methods, driven by the growing complexity, nonlinearity, and non-stationarity of real-world sequential data. Section 2 will discuss prior research on the proposed hybrid GAN-NLP framework, with a particular focus on statistical foundations, deep learning networks for sequential prediction, GANs, and the integration of textual data using NLP.

### 2.1 Statistical and Machine Learning Approaches to Time-Series Prediction

Classical statistical models have been used as basic tools for time series forecasting for many years. Among them, ARIMA models continue to be used widely because of their interpretability and well-known theoretical properties (Billings and Yang, 2006). More extensions allow time-varying volatility, e.g. Autoregressive Conditional Heteroskedasticity (ARCH) models, to better describe data when volatility is clustered (Engle, 1982). Although these models are useful as benchmarks, they are based on the assumptions of linearity and stationarity, which in many cases make them ineffective in complex, noisy settings.

The shortcomings of statistical models facilitated the adoption of machine learning methods capable of modelling nonlinear dependencies. Early machine learning methods, such as support vector machines (SVMs) and ensemble-based methods, proved more flexible and accurate than purely statistical models (Leung et al., 2014). Nevertheless, these methods usually require extensive feature engineering and are not easily scalable to high-dimensional sequential data.

### 2.2 Deep Learning for Sequential Modelling

Deep learning has greatly improved time-series modelling by enabling automatic representation learning from raw time-series data. RNNs, especially Long Short-Term Memory (LSTM), were developed to overcome the vanishing gradient phenomenon and learn long-term temporal dependencies (Hochreiter and Schmidhuber, 1997; Bengio et al., 1994). LSTM-based models have been used to good effect for a variety of sequential prediction tasks, including financial time series forecasting (often outperforming classical statistical baselines) (Fischer and Krauss, 2018; Nelson et al., 2017; Julian et al., 2023).

Even though LSTM models are successful, they are inherently discriminative and optimised for point estimation. This can lead to minimal distributional properties of data, especially in highly volatile or regime-switching settings. This limitation has led to a resurgence of interest in generative models that learn the underlying data distribution rather than being conditionally predictive.

### 2.3 Generative Adversarial Networks (GANs) for Time-Series Data

The innovations of Goodfellow et al. (2014) - referred to as Generative Adversarial Networks (GANs) - are a powerful framework for generative modelling, based on adversarial training between a discriminator and a generator. GANs have been incredibly effective in "learning" high-dimensional distributions with complex structures and have been studied especially in sequential and time-series data (Goodfellow, 2016).

A preliminary time-series forecasting model based on GANs showed that adversarial learning can yield a more robust and generalised predictor with a strong ability to model temporal dynamics and noise characteristics (Zhang et al., 2019). Recent works have since optimised GAN-based algorithms for sequential prediction and for time- and architecture-specific loss functions (Lazanas and Kampouropoulos, 2026). Specifically, Fin-GAN, where loss functions are driven by economic concerns to improve the forecasting and categorisation of financial time series, which is more distributionally realistic and directionally accurate (Vuletić et al., 2023). Extensive surveys demonstrate the growing applicability of GANs in time-series modelling, particularly in non-stationary and noisy settings (Brophy et al., 2023).

### 2.4 Natural Language Processing and Sentiment-Aware Prediction

The emergence of numerical time-series modelling and the increased availability of unstructured textual data have prompted investigations into how Natural Language Processing (NLP) techniques can be adapted for predictive systems. The contextual and sentiment-based information encoded in textual content on news and social media platforms cannot be measured directly and thus can influence sequential dynamics (Turney, 2002).

Empirical research has established that social media sentiment can be correlated with, and in certain instances anticipate, market trends. Applied research has also shown that aggregated sentiment measures derived from online content are associated with future shifts in market indices (Bollen et al., 2011). Subsequent studies found that sentiment cues offer predictive behaviour of time-series forecasts, particularly when using numerical characteristics (Derakhshan and Beigy, 2019; Puh and Bagić Babac, 2023).

### 2.5 Hybrid Deep Learning Models Integrating GANs and NLP

Recent studies have begun investigating hybrid architectures that combine deep learning with NLP-derived features to boost predictive performance. Text-based sentiment representations have been demonstrated to be effective in models that combine recurrent neural networks with purely numerical methods (Shen and Shafiq, 2020; Sonkiya et al., 2021). Nevertheless, numerous existing techniques rely on superficial fusion approaches or separately handle textual and numerical data streams.

More text-guided forecasting studies further confirm the importance of directly including textual information (e.g., news messages and channel descriptions) and learning cross-modal interactions through attention-based fusion. Xu et al. (2024) formalise Text-Guided Time Series Forecasting and demonstrate improved performance when combining textual cues with time-series representations via cross-attention mechanisms. Similarly, Emami et al. (2023) introduce a modality-conscious Transformer that utilises categorical text alongside numerical time series, employing feature-level and inter-modal attention, and emphasise the significance of organised multimodal fusion for forecasting. Lastly, language models are becoming increasingly crucial to modern NLP pipelines for deriving sentiment embeddings from longer-form financial text (e.g., analyst reports), which can serve as informative auxiliary predictors of price trends (Moreno and Ordieres-Mere, 2025). Other areas of application of hybrid modelling paradigms, including combining quantum circuits with deep neural networks for classification (Radhi et al., 2025), have also been studied recently, reflecting the wider trend of heterogeneous learning components in more complex predictive models.

Although time-series forecasting, generative modelling, and sentiment-aware prediction have made tremendous advances, the current literature tends to focus on them separately or on only a few integration methods (Li, 2025). Specifically, GAN-based methods emphasise distributional modelling of numerical sequences, whereas NLP-based methods use textual data without fully leveraging its interaction with temporal dynamics. In addition, most hybrid methods rely on superficial fusion rather than incorporating contextual information into the learning process. The limitations indicate the absence of coherent frameworks that integrate adversarial generative learning with sentiment-conditioned representations for time-series prediction in non-stationary settings.

## 3 Problem Formulation and Proposed GAN–NLP Framework

Our work aims to create a predictive framework for stock market data that can operate under non-stationary, noisy, and volatile conditions while leveraging contextual information from unstructured text. Historical numerical patterns alone influence stock prices, but exogenous information, including market sentiment and collective perception, cannot be directly reflected by numerical indicators. The purpose of the proposed approach is therefore to combine numerical stock market data and sentiment-based contextual information into a single learning model.

Let:$\boldsymbol{x_t} \in \mathbb{R}^d$, denote a multivariate numerical observation of a financial asset at trading day $\boldsymbol{t}$, where $\boldsymbol{d}$ corresponds to the number of numerical attributes describing the stock. In this study, these attributes include price-related and trading-related variables such as: {*Open, High, Low, Close, Adjusted Close*, and *Volume}*, as summarised in **Table 1**:

**Table 1** Numerical stock market attributes used as model inputs and their role in price prediction

| Attribute | Description | Functionality in Prediction |
|---|---|---|
| *Open* | The opening price of the stock on a specific day | Indicates investor sentiment at the start of trading. A higher-than-usual open might suggest buying pressure and a potential price rise. |
| *High* | The highest price the stock reached during the day | Indicates buying pressure and potential upside. Resistance levels can form around previous highs, making them difficult to break through. |
| *Low* | The lowest price the stock reached during the day | Indicates selling pressure and potential downside. Support levels can be identified around previous lows, which may be difficult to fall below. |
| *Close* | The price at which the stock last traded on a specific day | Often considered the most important price point, as it reflects the final sentiment of the day's trading. |
| *Adjusted Close* | The closing price adjusted for stock splits and dividends | Provides a more accurate picture of price movements over time, especially when comparing periods with corporate actions. |
| *Volume* | The number of shares traded on a specific day | High volume can indicate increased interest in the stock, potentially leading to higher volatility and price changes. Low volume can suggest a lack of investor interest. |

Given a historical window of length $L$, $X_t = \{x_{t-L+1}, \dots, x_t\}$, the prediction task consists of estimating the next-day stock observation $x_{t+1}$ (or a short prediction horizon) conditioned on historical market behaviour. Simultaneously, unstructured text information matched to the trading period is processed to generate sentiment-based representations. In fact, the numerical input $X_t$ is constructed as a moving multivariate window of historical stock market observations, with each component $x_t$ associated with a price- and trading-related set of attributes, as outlined in Table 1. The window moves chronologically through trading history, preserving the time sequence of market data and enabling the model to capture short- and medium-term correlations in stock behaviour. This representation allows the learning model to utilise the joint dynamics of many numerical attributes, without requiring the underlying market process to be stationary or linear.

Also, let: $s_t \in \mathbb{R}^k$, represent a feature vector of sentiment obtained by applying Natural Language Processing methods, such as sentiment analysis to social media posts. Such characteristics encode contextual data, related to market perception and investor behaviour, which can affect price formation but cannot be observed directly in numerical stock indicators. The learning objective is therefore formulated as modelling the conditional distribution: $p\left( x_{\{t+1\}} \mid x_{\{t-L+1\}}, \dots, x_t, s_t \right)$, instead of working on point-wise prediction error only. In the suggested framework, the sentiment vector $s_t$ is treated as a built-in auxiliary conditioning signal rather than as an independent predictive input. This design option allows the learning process to augment numerical market dynamics with contextual information and remains robust when textual information becomes sparse, noisy, or weakly informative.

The following formulation encourages the use of generative modelling better suited to capturing uncertainty, regime changes, and the distributional characteristics typically encountered in financial markets. In this regard, the proposed framework utilises a Generative Adversarial Network (GAN), in which a generator G is trained to generate realistic future stock observations from historical numerical data and contextual features derived from sentiment analysis.

On a high level $G$ applies a mapping of the form: $\hat{x}_{t+1} = G(X_t, s_t)$, in contrast to a discriminator $D$, which looks at the plausibility of an observation $x$ by estimating: $D(x, X_t, s_t) \in [0, 1]$, indicating whether the input corresponds to a real or generated sample conditioned on historical market behaviour and sentiment information. Through adversarial feedback from $D$, the generator progressively improves its ability to produce context-aware and temporally coherent stock predictions. This concept enables the model to collectively utilise time-market and situational sentiment information, without the constraining assumptions of linearity and stationarity in classical statistical models. Moreover, the framework can be designed to incorporate sentiment information as an auxiliary conditioning signal rather than an independent predictor, making it resilient when textual signals are scarce, noisy, or poorly informative. Figure 1 offers a conceptual description of the above and provides an illustration of the main aspects of our hybrid $G/D/s_t$ (Generator – Discriminator – Sentiment) interaction on which the training of the GAN is based.

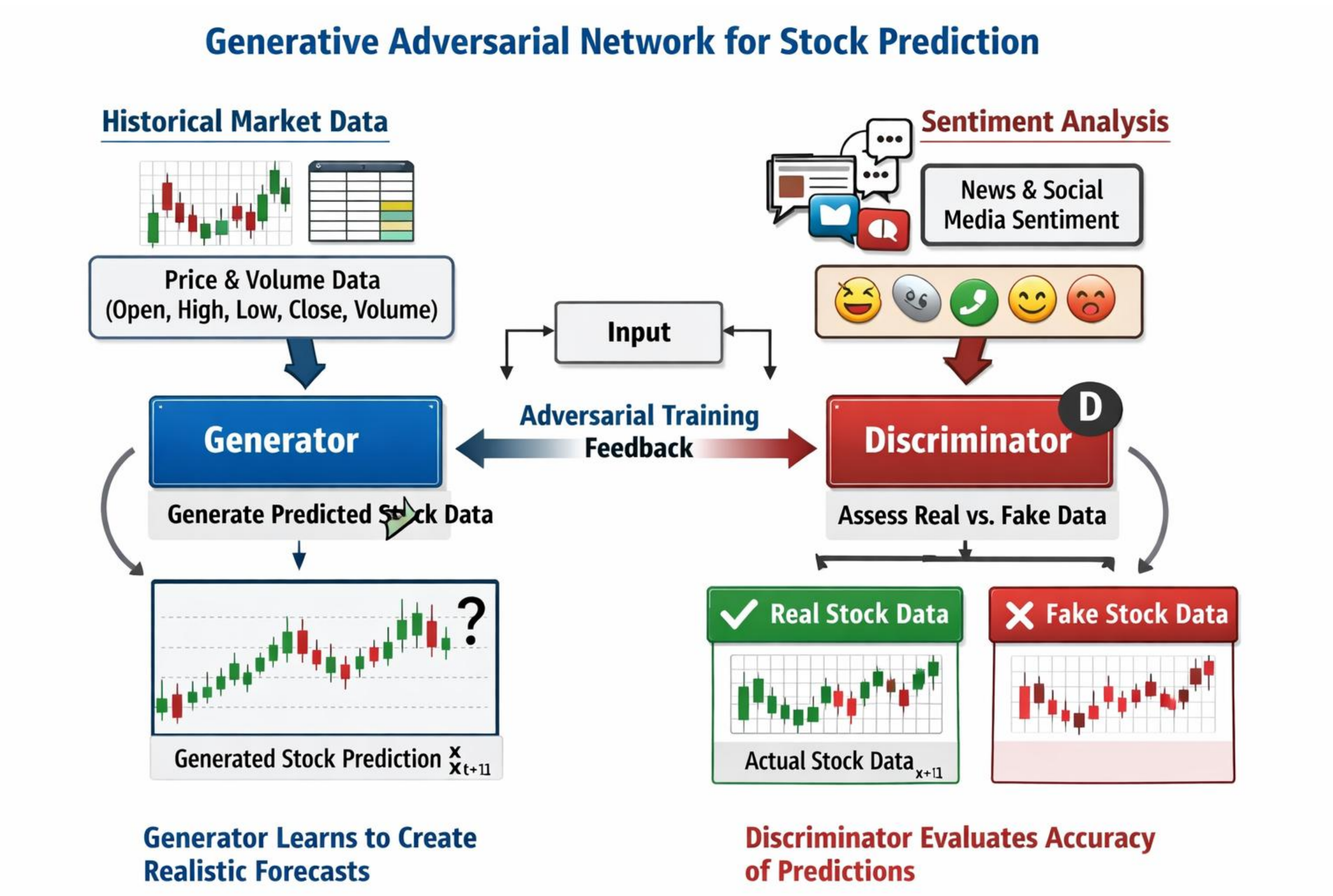

**Fig. 1** The composite formulation of our hybrid GAN model allows the model to share and collectively explore temporal dependencies, contextual sentiment cues and adversarial feedback during training

The general training is based on an adversarial learning process, with the generator trained to produce realistic future observations from past numerical data and sentiment features, and the discriminator assessing the consistency of predictions with actual market behaviour. The two components are updated during training, and the generator is gradually enhanced to better capture both the temporal and contextual signals. The arrangement allows the model to acquire joint market dynamics and sentiment information representation in a single framework.

## 4 Experimental Evaluation

### 4.1 Data Sources and Preprocessing

The assessment comprises large-cap U.S. equities, which are typically associated with high liquidity and strong market effects. More precisely, the discussion focuses on seven representative stocks: *Apple (AAPL), Microsoft (MSFT), Amazon (AMZN), Alphabet (GOOGL), Meta Platforms (META), Tesla (TSLA), and Nvidia (NVDA)*. The evaluation design relies on numerical market data and textual sentiment information, both described below:

1) **Numerical market data:** The stock market data are obtained daily through Yahoo Finance[1] and are composed of the attributes described in Table 1 (Section 3). The information is sorted chronologically and includes only trading days. Redundant records are eliminated, and missing values are looked into and processed to maintain temporal continuity. Feature selection is used to retain the variables associated with price and trading that are important for prediction. Before model training, numerical features are Min-Max scaled, and the scaling parameters are estimated only on the training subset and then applied to the validation and test data to avoid information leakage.

[1] https://finance.yahoo.com/

2) **Textual and sentiment data:** The textual data for the stocks under study are gathered from posts on social media (tweets) retrieved through the Twitter (X) platform[2]. Sentiment is then extracted from the raw text using "cleaning" procedures, which include normalisation and noise elimination. Sentiment analysis is performed using the VADER (Valence Aware Dictionary and sEntiment Reasoner) lexicon-based approach, which stores only the sentiment scores of the compounds and matches them to the corresponding trading days. In cases where there exist several text entries for the same day, these are combined into one daily sentiment representation to allow compatibility with the daily frequency of the numerical data. VADER is driven by its rule-based nature, which offers consistent and interpretable sentiment scores for short and noisy financial text, including social media posts. VADER enables uniform sentiment extraction without fine-tuning on domain-specific data, unlike transformer-based models, which require large annotated corpora and add extra training complexity. This decision makes it possible to attribute the differences in predictive performance evident in this study to the modelling framework rather than to the variability introduced by advanced language models. However, the combination of transformer-based sentiment representations is also an encouraging avenue for future research.

All preprocessing procedures are applied uniformly across assets and evaluation settings to ensure reproducibility and fair comparison.

### 4.2 Evaluation Protocol

The testing of the suggested framework is presented here as a systematic process that outlines the modelling and evaluation decisions made. The primary points of the assessment plan that uses two base models (ARIMA and LSTM) compared to our approach are outlined below:

1. **Temporal data partitioning**: In all models, the observations of the stock prices are arranged in a chronological order to maintain causality in time. There is no random shuffling at any point of assessment. Different out-of-sample partitioning strategies are used depending on model type, as discussed below, as well as on the different modelling needs and evaluation practices of each approach.
2. **Model-specific data splits**: **a)** in the case of the ARIMA baseline, around 90% of available observations are utilized in fitting the model, and the remaining 10%, which is the latest portion of the time series, is set aside to be used in out-of-sample testing, **b)** in the case of the LSTM model, the time-varying data is separated into training and test data in the ratio of 70% and 30%, respectively and **c)** in our GAN-based model, the final 20 observations of each time series are kept to make the evaluation, and the rest of the data are utilized to train the model. It is a decision that allows the generated and observed values to be directly compared over a predetermined prediction period. These partitioning methods are used to ensure that any evaluation is conducted strictly out-of-sample and reflects realistic forecasting conditions. The adoption of a model-specific temporal partitioning scheme is methodologically motivated, as different modelling paradigms (statistical, recurrent, and adversarial) have different assumptions regarding data availability, convergence behaviour, and training stability, and using a uniform split would hurt some models or deviate from their standard evaluation practices. The specific partitioning schemes (ARIMA: 90%–10%, LSTM: 70%–30%, GAN: last 20 observations) are selected to reflect the distinct training and evaluation requirements of each modelling paradigm, while preserving strictly out-of-sample testing conditions.
3. **Prediction task definition:** Evaluation of all models is done on a one-step-ahead forecasting task. For each trading day $t$, it is based on available information (up to time) to generate predictions of the next trading day. The models maintain the same prediction horizon so that the predictive accuracy can be compared directly.

---

[2] https://x.com/

4. **Models under evaluation**: The analysis considers three types of models, namely:
    a. A statistical ARIMA-based baseline, which incorporates linear time-dependencies in stock prices (Billings and Yang, 2006). The parameters of ARIMA models are chosen by mixing an information criterion and automated order selection and stationarity tests.
    b. A Long Short-Term Memory (LSTM) network-based deep learning baseline, which aims to learn the nonlinear sequential behaviour of numerical stock data.
    c. Our suggested GAN-based model with sentiment conditioning the framework, which approximates the conditional distribution of future stocks' prices based on historical numerical data and contextual sentiment analysis. Each model is tested using temporally aligned data from the same stock "universe".
5. **Model training configuration**: The models based on neural networks are trained with fixed configurations that are aligned with the original study. In the LSTM model, early stopping and learning-rate scheduling are used during training to enhance convergence stability (Kim et al., 2021; Prechelt, 2012); thus, in the GAN-based model, the numerical price values are scaled to the range ***(-1, 1)***. The training is done with a fixed batch size of 5, and the prediction period is one trading day. Scaling parameters are obtained from the training data and reused during evaluation.
6. **Evaluation metrics**: The predictive performance is determined through standard regression-based measures that are widely used in stock price forecasting, such as:
    a. Mean Absolute Error (MAE)
    b. Mean Squared Error (MSE)
    c. Root Mean Squared Error (RMSE)
    d. Mean Absolute Percentage Error (MAPE), where applicable.

    All metrics are calculated only on held-out evaluation data and reported uniformly across stocks and models.

Figure 2 summarises the evaluation workflow adopted in this study.

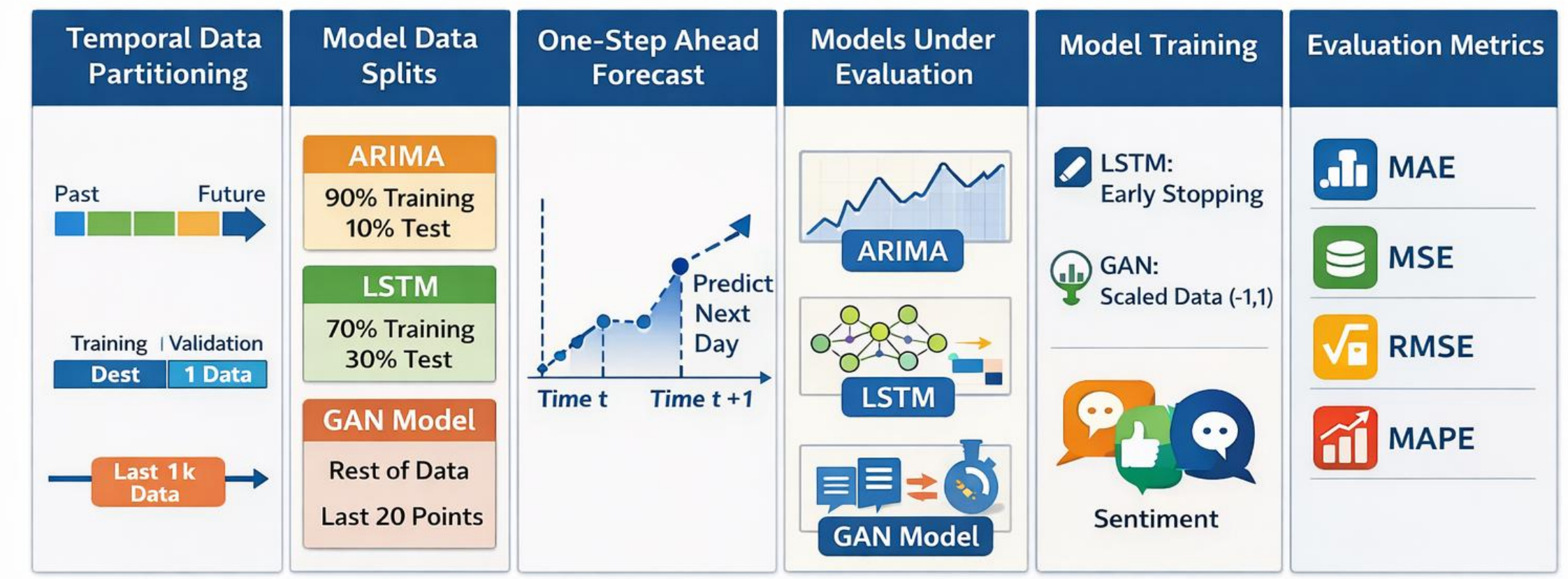

**Fig. 2** Overview of the evaluation workflow adopted in this study, illustrating data preprocessing, temporal partitioning, model assessment, and performance evaluation

The main training and configuration parameters used for the LSTM and GAN-based models are summarised in Table 2. In implementing the proposed framework, both the generator and discriminator are constructed as fully connected neural networks. The generator consists of multiple hidden layers with ***ReLU*** activation functions, while the discriminator employs ***LeakyReLU*** activations to improve gradient flow during adversarial training. The models are trained using the Adam optimiser, and the adversarial objective follows the standard minimax GAN formulation. Training is performed over a fixed number of epochs, with alternating updates between the generator and discriminator to ensure stable convergence.

**Table 2** Model Training Parameters

| Parameter | LSTM Model | GAN Model |
|---|---|---|
| Optimizer | Adam | Adam |
| Learning Rate | 0.001 | 0.0002 |
| Batch Size | 32 | 5 |
| Input Scaling | Min–Max | (−1, 1) normalization |
| Training Strategy | Supervised | Adversarial training |
| Early Stopping | Yes | No |
| Epochs | Fixed / adaptive | Fixed |

## 5 Results and Performance Analysis

This Section presents the empirical findings derived from the evaluation protocol in Section 4. All evaluation tests share the same preprocessing steps, temporal data partitioning strategies and one-step-ahead forecasting setup to ensure strictly out-of-sample evaluation and comparison between models and assets. The analysis is conducted consistently across the seven large-cap U.S. equities examined, using the same evaluation metrics.

The proposed sentiment-conditioned GAN framework is compared with two baselines: an ARIMA model and an LSTM model. Predictive performance is evaluated only on held-out test data using standard regression measures (MAE, MSE, RMSE, and MAPE, where applicable), enabling systematic comparison of forecasting performance and robustness under realistic market conditions.

### 5.1 Comparative Forecasting Performance

In this subsection, a comparative analysis of the forecasting performance of the three modelling methods considered in this paper, i.e., the ARIMA baseline, the LSTM-based deep learning model, and our sentiment-conditioned GAN modelling framework, is presented. Table 3 presents the predictive performance of all the models based on the conventional regression statistics, which are MAE, MSE, RMSE, and MAPE, where applicable. The findings suggest that none of the models performs equally well across all assets, but our sentiment-conditioned GAN model shows consistently high performance in several instances, especially among stocks with higher volatility and stronger sentiment-driven dynamics. Here, the inclusion of sentiment information seems to improve the model's ability to capture short-term market dynamics that would otherwise be impossible to capture with pure numbers.

The LSTM baseline is competitive for stocks with a more predictable time-frequency structure and, in many cases, outperforms the statistical ARIMA model in terms of error size. Although ARIMA is an appropriate model for linear dependencies, it tends to have larger prediction errors than neural methods, especially during periods of high market volatility.

**Table 3** Comparative Forecasting Performance

| Stock | Model | MAE[3] | RMSE | MSE | MAPE |
|---|---|---|---|---|---|
| Google | ARIMA | 14.04 | 16.62 | 276.17 | 0.11 |
| Google | LSTM | - | 6.97 | 48.58 | - |
| Google | GAN | - | 13.42 | 180.10 | - |
| Amazon | ARIMA | 16.83 | 20.87 | 435.67 | 0.13 |

[3] For LSTM and GAN models, MSE values are derived from the reported RMSE values using $MSE = RMSE^2$. MAE values are not reported for these models because the original evaluation pipeline retained RMSE as the primary comparable error metric.

| Amazon | LSTM | - | 3.35 | 11.22 | - |
|---|---|---|---|---|---|
| Amazon | GAN | - | 7.05 | 49.70 | - |
| Apple | ARIMA | 9.38 | 11.83 | 139.97 | 0.05 |
| Apple | LSTM | - | 6.24 | 38.94 | - |
| Apple | GAN | - | 7.02 | 49.28 | - |
| Meta | ARIMA | 128.89 | 146.11 | 21350.1 | 0.47 |
| Meta | LSTM | - | 11.21 | 125.66 | - |
| Meta | GAN | - | 8.24 | 67.90 | - |
| Microsoft | ARIMA | 34.75 | 41.22 | 1698.81 | 0.10 |
| Microsoft | LSTM | - | 14.76 | 217.86 | - |
| Microsoft | GAN | - | 27.07 | 732.78 | - |
| Nvidia | ARIMA | 155.24 | 182.04 | 33139.52 | 0.39 |
| Nvidia | LSTM | - | 118.30 | 13994.89 | - |
| Nvidia | GAN |  | 13.39 | 179.29 |  |
| Tesla | ARIMA | 25.30 | 30.70 | 942.59 | 0.13 |
| Tesla | LSTM | - | 13.21 | 174.50 | - |
| Tesla | GAN | - | 9.33 | 87.05 | - |

As shown, the findings in Table 3 demonstrate that our sentiment-conditioned GAN model can provide quantifiable benefits to prediction accuracy in realistic market behaviour, and point to the complementary abilities of LSTM/RNNs to predict assets with smoother price behaviour.

**5.2 Qualitative Forecasting Comparison**

To further enhance the quantitative analysis of Section 5.1, this subsection provides a detailed qualitative analysis of representative predicted-vs.-actual price paths. The objective is not only to visualise the accuracy of the forecasts, but to understand how various modelling paradigms react to various market features, which may be the volatility, persistence of the trend, or the sensitivity to the information, which is driven by external sentiment-driven factors. Each figure refers to held-out test data and represents purely out-of-sample behaviour.

*Meta* is a representative of stocks with high volatility and dynamics driven mainly by sentiment. As Figure 3 demonstrates, the predictions made by our sentiment-conditioned GAN model are better at tracking sudden price fluctuations and short-term directional shifts. Specifically, the model adjusts more quickly to drastic changes in the price series, minimising lag effects, which are common in purely sequential models. This behaviour implies that sentiment conditioning is an informative contextual signal that enables the generative process to predict market responses of external events, news cycles, or changes in investor perception that are not directly encoded in historical prices.

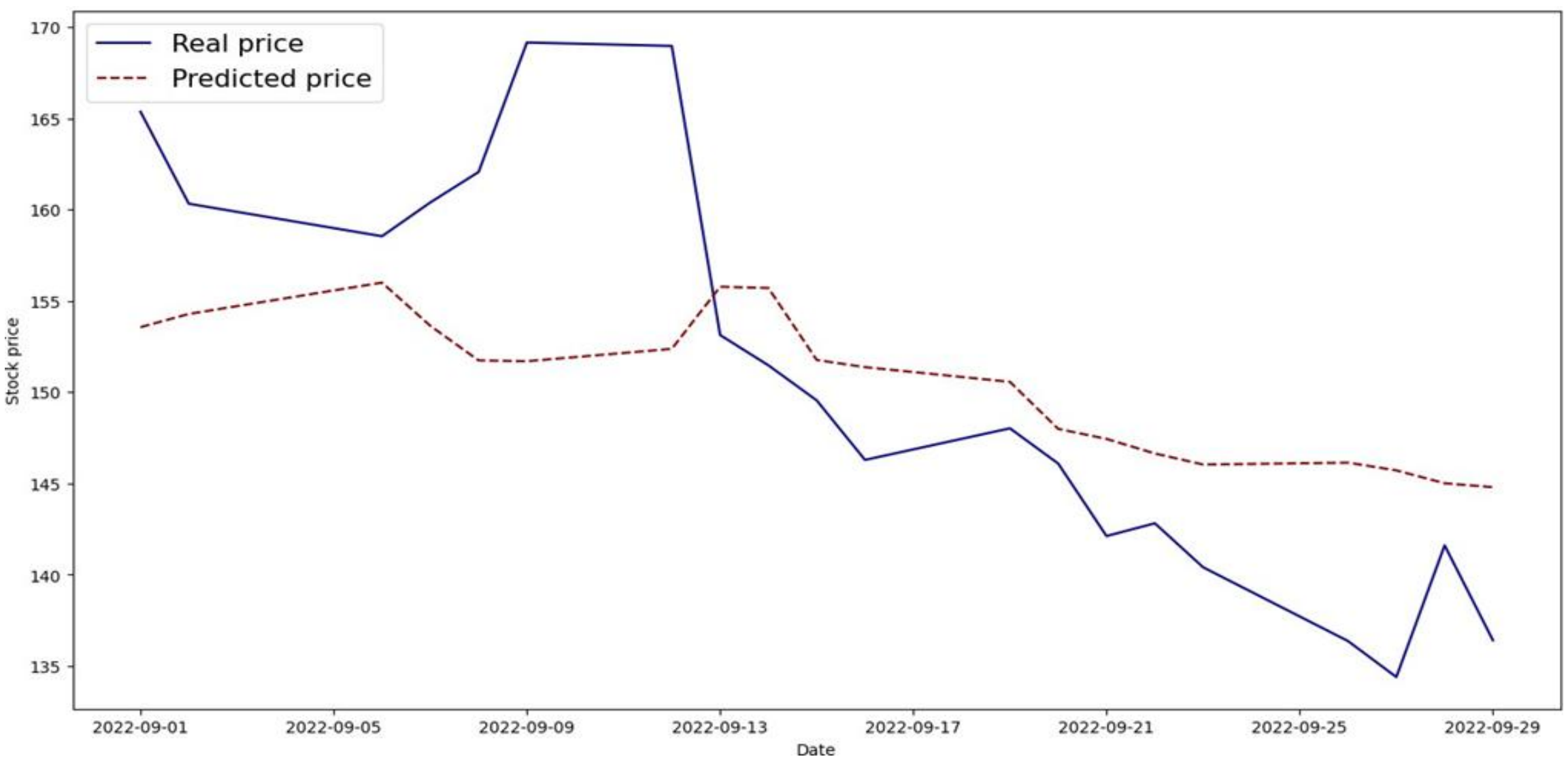

**Fig. 3** Predicted versus actual closing prices for Meta (*META*) on held-out test data using our sentiment-conditioned GAN model

Conversely, assets with more continuous price dynamics and more stable temporal patterns reveal weaknesses in our sentiment-conditioned GAN model. Figure 4 shows this behaviour by presenting the predicted vs actual price behaviour of a relatively stable asset, such as Alphabet Inc. (Google), using the GAN-based model. The forecasts generated in this case exhibit greater variability relative to the actual price path, particularly in their ability to capture slow trends and long-term directional movements. The qualitative mismatch is a sign that when market dynamics are largely determined by the laws of history rather than sudden sentiment signals, the GAN process is unreliable at maintaining correct long-term alignment. In line with the quantitative findings in Section 5.1, the LSTM-based model appears to perform better in these scenarios due to its recurrent nature, which is better suited to learning long-term temporal dependencies.

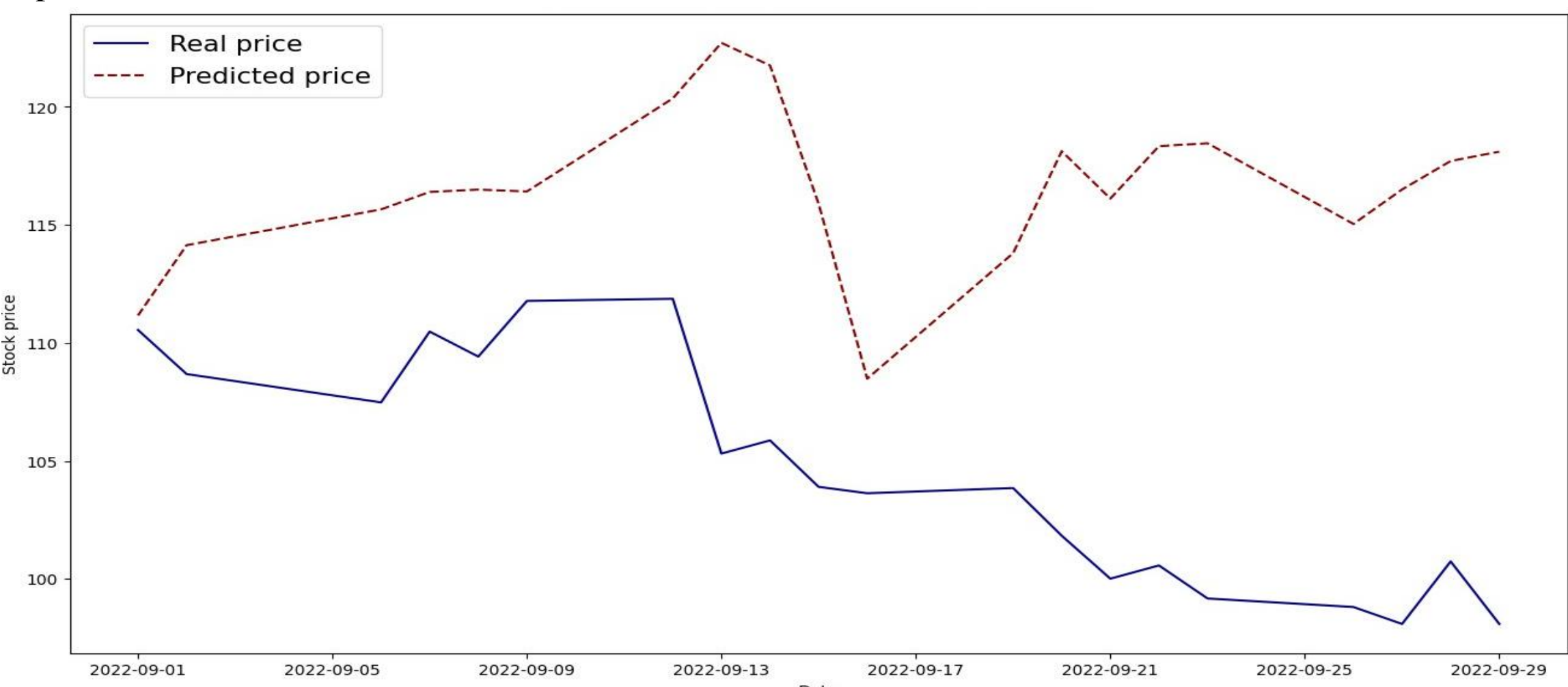

**Fig. 4** Predicted vs. actual closing prices for Alphabet Inc. (*Google*) with our GAN modelling framework, illustrating reduced forecasting accuracy for assets with smoother temporal dynamics

Another example that could be used to demonstrate the complementary qualities of the analysed models is Tesla (*TSLA*). Tesla presents a challenging forecasting problem due to its volatility and price fluctuations driven by announcements, social media, and speculative activity. Our sentiment-conditioned GAN model, shown in Figure 5, shows a higher reaction to abrupt shifts in price direction than the sequential-only approaches. Although certain deviations are unavoidable given the asset's unpredictability, the generative model adapts its response to sharp changes, which supports the hypothesis that sentiment-sensitive modelling can help achieve greater adaptability in highly dynamic market environments.

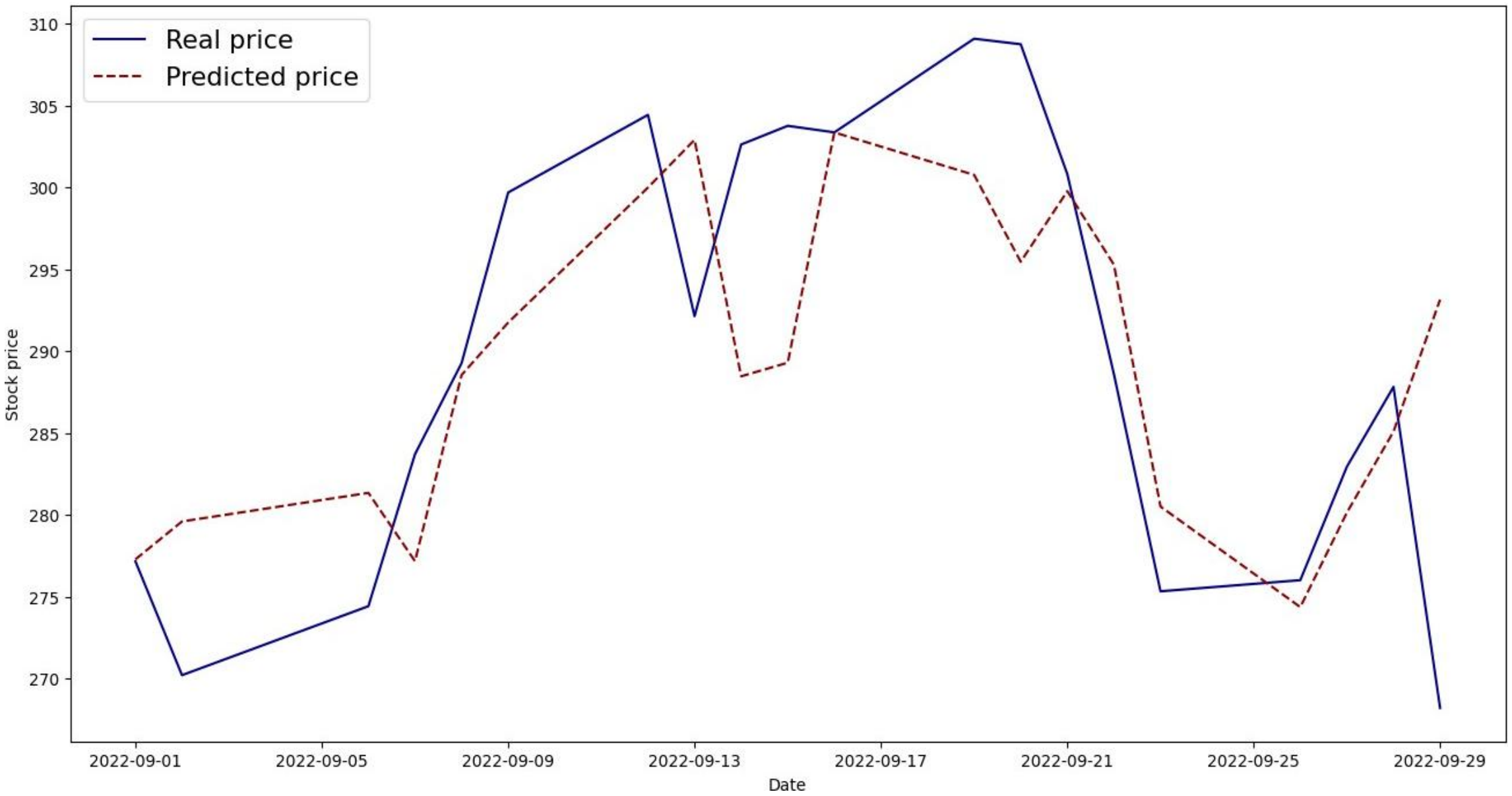

**Fig. 5** Predicted vs. actual closing prices for Tesla (TSLA) with our GAN modelling framework; we notice the model's ability to capture abrupt price movements in a highly volatile, sentiment-driven market

Taken together with the qualitative evidence, the quantitative results in Section 5.1 are strengthened. The graphical comparisons affirm that there is no universal best modelling method that fits all assets, but that model effectiveness is highly dependent on the underlying market regime. Recurrent neural networks like LSTMs can be used to model assets with predictable, time-dependent variations, and our sentiment-conditioned GAN architecture offers specific benefits in environments where price dynamics are significantly influenced by exogenous information and investor sentiment. The observations highlight the significance of model selection and a hybrid modelling approach in practical stock price forecasting.

### 5.3 Aggregate Robustness and Comparative Performance Analysis

Building on the earlier per-asset-based analysis detailed in the preceding subsections, this section provides a more aggregate, robustness-based analysis of the forecasting performance of all the assets discussed. Although stock-specific measures can provide a comprehensive understanding of individual behaviour, aggregate measures are needed to assess the overall consistency, stability, and relative effectiveness of the models tested in the heterogeneous market.

Table 4 presents aggregate performance statistics based on the RMSE values reported in Section 4, including mean RMSE, median RMSE, and the "number of wins" (i.e., the number of assets for which each model produces the lowest prediction error). A combination of mean and median RMSE enables evaluation that considers overall accuracy and minimises the impact of extreme error values from highly volatile assets.

**Table 4** Aggregated Robustness Metrics Across Models

| Model | Mean RMSE | Median RMSE | Wins (Lowest RMSE) |
|---|---|---|---|
| **ARIMA** | 64.2 | 30.7 | **0** |
| **LSTM** | 24.86 | 11.21 | **4** |

| **GAN** | 12.02 | 8.79 | **3** |
|---|---|---|---|

Although the robustness analysis is mainly based on aggregate statistics (RMSE), other assessment metrics can also be used to better understand practical forecasting performance, such as directional accuracy or risk-adjusted performance measures. These extensions can be viewed as directions for future work. In particular, trading simulation or back testing frameworks could provide a more application-oriented assessment of model performance, although such analysis is beyond the scope of the present study.

The cumulative findings reveal that the LSTM-based model yields the lowest mean and median RMSE across assets, a strong indicator of overall consistency in capturing the temporal relationships among stock price fluctuations. Conversely, the sentiment-conditioned GAN model has fewer total wins but competes on aggregate terms, underscoring its usefulness for a subset of assets with greater volatility and sentiment sensitivity. The ARIMA baseline shows significantly higher aggregate errors, underscoring its weaknesses in analysing complex, non-linear market dynamics.

In general, the results of our evaluation support the asset-conditional character of model performance and provide a robustness-oriented summary that supplements the more detailed analysis of stock levels discussed above. These aggregate findings are a quantitative basis for the comparative and qualitative discourse of the next Section.

## 6 Discussion and Future Work Directions

The selection of ARIMA and LSTM as baseline models is founded on their extensive application in time-series forecasting. This provides a means for comparative analysis. Some researchers have explored methodologies such as GARCH-based volatility models or Transformer-based architectures. In this study, we aim to evaluate the effectiveness of generative modelling when integrated with sentiment conditioning. We do not intend to examine every forecasting method currently available. Future investigations could consider incorporating additional baseline models, a prudent direction for further research. The empirical assessment shows that the performance of stock price forecasts is always conditional on asset-specific factors and current market regimes, rather than governed by a single modelling paradigm. This finding is in line with recent insights into machine learning-based financial forecasting, which highlight the importance of adaptive modelling techniques over general predictors (Zhang et al., 2024). Results reported in this context are not to be interpreted as a comparison of architectures, but rather as indicators of how various learning mechanisms respond to different informational environments. Sequential deep learning structures are highly effective in environments where price dynamics follow quite stable time series. Recurrent memory models are best suited to such dependencies, having been initially developed in the original work on LSTM networks (Hochreiter and Schmidhuber, 1997) and since then applied to financial forecasting tasks (Nelson et al., 2017; Fischer and Krauss, 2018). Historical price data seem adequate for producing accurate short-term forecasts when market forces are largely endogenous, thereby limiting the marginal value of external contextual information.

By contrast, incorporating sentiment information is most effective for stocks whose price action is more susceptible to external stories, news, cycles, and investor perceptions. Previous research has demonstrated that sentiment measures derived from textual data can capture aspects of market psychology not directly reflected in quantitative price data (Bollen et al., 2011; Azar, 2016; Derakhshan and Beigy, 2019). Training generative models with these cues enables the prediction procedure to respond to sudden informational shocks, especially in a high-volatility setting. Notably, the selectivity of this advantage supports the view that sentiment is a regime-specific indicator rather than a universally predictive one. These results have broader implications for multimodal data fusion in financial forecasting. The literature has shown that naive fusion approaches, including naive feature concatenation, do not typically fully leverage the complementary characteristics of heterogeneous data sources (Puh and Bagić Babac, 2023;

Moreno and Ordieres-Mere, 2025). Our findings indicate that conditioning processes, in which contextual information actively determines the generative or predictive process, are better frameworks for capturing distributional uncertainty. This is in line with the current developments of conditional generative modelling and sentiment-directed forecasting (Sonkiya et al., 2021; Vuletić et al., 2023).

The relatively poor results of traditional statistical methods underscore the inadequacy of linear assumptions in modern financial markets. Although autoregressive models remain applicable for interpretability and benchmarking (Box et al., 2015), they have representational weaknesses and can be designed to address nonlinear dependence, regime change, and exogenous information flow. Other recent comparative surveys have reported similar findings, highlighting the widening gap in the performance of classical and deep learning architectures in complex market settings (Rouf et al., 2021; Zhang et al., 2024; Benidis et al., 2022). Our findings also have wider implications for the general debate on market behaviour and information efficiency beyond predictive considerations. The fact that sentiment-conditioned models have performed better in certain situations aligns with theories of behavioural finance, which highlight the significance of investor sentiment and limited rationality in the dynamic of short-term price movements (Kahneman and Tversky, 1979; Shiller, 2015). At the same time, these findings do not imply constant inefficiencies or predictability since short-lived informational impacts can quickly fade as markets adapt. This reading is consistent with a moderate opinion of market efficiency, in which temporary aberrations can be compatible with long-term adaptive behaviour (Fama, 1970).

The evaluation in this study focuses on one-step-ahead forecasting, which allows a controlled comparison of model behaviour under identical prediction horizons. While multi-step or rolling-window forecasting could provide additional insights into long-term stability, the selected setup is consistent with standard evaluation practices in financial time-series prediction and enables direct comparison across modelling paradigms. The extension to multi-horizon forecasting is considered a relevant direction for future research. There are some shortcomings to consider when explaining the current findings. The data are analysed using daily information and sentiment derived from social media, which may not capture intraday changes or longer-term information impacts. Previous research has indicated that sentiment signals are noisy and platform-specific and could be biased unless put into context (Turney and Littman, 2003; Whissell, 1989). Further, even sophisticated architectures struggle to capture extreme volatility, highlighting the uncertainty of financial markets and the challenge of maintaining stable predictive performance in highly turbulent markets. The comparative evaluation is based on deterministic error metrics (MAE, MSE, RMSE) computed on strictly out-of-sample predictions. While statistical significance testing, such as the Diebold–Mariano test, could provide additional insights into the relative predictive accuracy of competing models, the focus of this work is on comparative performance trends and robustness across different assets and modelling paradigms. The inclusion of formal statistical testing constitutes a valuable direction for future work.

Future research can expand on this study by investigating regime-conscious architectures that dynamically adjust their reliance on numerical and textual inputs. Another attractive direction is the integration of alternative contextual sources, e.g., macroeconomic indicators, analyst reports, or option-implied measures (Xu et al., 2024; Woo et al., 2025). Furthermore, explainability methods may be included to make the implementation more transparent and practically relevant, especially in high-stakes financial implementations. Together, these guidelines demonstrate that context-sensitive deep learning models can be used to make financial predictions while accounting for the structural uncertainty in market dynamics.

## Conflicts of Interest

All authors declare that they have no conflicts of interest.